\documentclass[prd,aps,12pt]{revtex4}
\usepackage{graphicx}
\newcommand{\bb}{\begin{eqnarray}}
\newcommand{\ee}{\end{eqnarray}}
\begin{document}
\title{ \textbf{SCALAR AND SPINOR PARTICLES WITH LOW BINDING ENERGY IN THE STRONG STATIONARY
MAGNETIC FIELD STUDIED BY MEANS OF TWO- AND THREE-DIMENSIONAL MODELS
}}
\author{V.N. Rodionov}
\email{vnrodionov@mtu-net.ru} \affiliation{Physical Chair of Russian State Geological Prospecting
University 118816, Moscow, Russia}

\begin{abstract}
On the basis of analytic solutions of Schrodinger and Pauli
equations for a uniform magnetic field and a single attractive
$\delta({\bf r})$-potential the equations for the  bound
one-active electron states are discussed. It is vary important
that ground electron states in the magnetic field essentially
different from the analog state of spin-$0$ particles that binding
energy has been intensively studied at more then forty years ago.
We show that binding energy equations for spin-$1/2$ particles can
be obtained without using  of a well-known language of boundary
conditions in the model of $\delta$-potential that has been
developed in pioneering works. Obtained equations are used for the
analytically calculation of the energy level displacements, which
demonstrate nonlinear dependencies on field intensities. It is
shown that in a case of the weak intensity a magnetic field indeed
plays a stabilizing role in considering systems. However the
strong magnetic field shows the opposite action. We are expected
that these properties can be of importance for real quantum
mechanical fermionic systems in two- and three-dimensional cases.
\end{abstract}

\pacs{PACS numbers: 03.65.-w, 03.65.Ge}

\maketitle

\section{  Formulation of the problem.}

The effect of an external electromagnetic field on nonrelativistic
charged particles systems (like atoms, ions and atomic nucleuses)
has being investigated systematically for a long time (see, for
example, ~\cite {Kel.1, Fr, Zel, Dem.1, Dem.2, Kel.2, Ser, Asp,
Rit, Druk, Nik, Baz, Zeld, Pop, Man} ). Though this problem has a
long history, a set of questions as before requires additional
studying. For example, till now the systematic analysis of the
bound states of particles with spin-1/2 in the intensive magnetic
field is absent. Note that the basic results for the case of
spinless particles were obtained by using of analytical solutions
in nonperturbative mathematical treatments. As usually the exact
solutions of Schr\"{o}dinger equations with Hamiltonians taking
into account an particle bound by short-range potential in the
presence of external fields are used. Furthermore there is  the
rather common opinion on invariably the stabilizing role of
magnetic field in decays of quasistationary states~\cite {Zel,
Pop, Man}. This point of view is caused by the fact that spinor
states of electrons in an external electromagnetic field are not
usually taken into account in nonrelativistic treatments that is
not always adequate ~\cite {RoKr1, RoKr2}. In this paper we treat
an essential part of these problems.

There are we consider  charged spin-$0$ and spin-$1/2$
particles  bounded by a short range
potential (a $\delta$-potential) and located in the external
stationary magnetic field with an arbitrary intensity.
Note that the potential of zero-radius is a widely spread
approximation for a multi-electronic atom field
and especially for a negative ion field~\cite {Baz, Dem.3}.

For the particle in a $\delta$-potential and in the magnetic field
one can see the energy level displacements. The most adequate
instrument for the investigation of such states is the development
of the binding energy equation formalism (\cite {Dem.1, Baz,
Lan}).

When an electron moves in a uniform magnetic field  oriented in
$z$-direction this quantum mechanical system is invariant with
respect to $z$-axis. Then the system  becomes essentially
two-dimensional in the $xy$ plane. Many physical phenomena
occurring in quantum systems of electrically charged fermions,
which have the axial symmetry, can be studied effectively by means
of the equations of motion in 2+1 dimensions - quantum Hall
effect, phenomenon of high-temperature superconductivity,
investigation of different film defects and etc. A number of this
effects in constant magnetic fields including the investigation of
a certain type of doped two-dimensional semimetals one can find
in~\cite{Sc},\cite{Ne} (treatment of two-dimensional models see
also in~\cite{chib} - \cite{Kh2}). However there are a lot of
physical phenomena which as before occur in three-dimensional
space. In this paper we investigate the effect of a stationary and
uniform magnetic field on localized electron states with regard of
$2+1$ and 3+1 dimensions.

The main purpose of this work is to derive equations for the
binding energy of an fermion in the field contained an attractive
singular potential and a stationary external magnetic field in
two- and three-dimensional cases. This treatment is realized by
means of the standard quantum mechanical methods using the
development of the unknown wave function  in a series at the
eigenfunctions obtained for the fermionic system in the pure
magnetic field. This formalism is principally different from the
traditional derivation of  wave functions in similar tasks by
using the boundary condition typical for the
$\delta$-potential~\cite {Dem.1,Dem.2, Pop, Man}. It is very
important that our approach let us to develop the consistent
investigation of the spin effects arising in the external magnetic
field.

The general structure of the paper and its main results we can
formulated as follows. In the second Section on the basis of
explicit solution for Schr\"{o}dinger equation the  equation for
scalar  particles with a low binding energy is constructed in an
external stationary magnetic field. In the third Section on the
basis of the analog analysis  of explicit solutions for Pauli
equations the expressions for the  energy of bound electron states
in the $\delta$-potential and in the external magnetic  fields are
obtained. Finally in the fourth Section the equations for the
bound energy of  spin-$0$ and spin-$1/2$ particles in the presence
of a weak and strong magnetic fields is simultaneously discussed
because early in similar tasks the spin of particles is not
adequate taken into account.

\section{A scalar particle in an attractive potential in the presence
of a uniform magnetic field}

Let us consider a charge  in
a uniform magnetic field ${\bf B}$, which is specified as
\bb
 {\bf B}=(0,\,0,\,B)=\nabla\times {\bf A},\ \
 {\bf A}=(-yB,\,0,\,0).
\label{e1}
 \ee
 The Schr\"{o}dinger equation in field (\ref{e1}) has the form

\bb
 i\hbar\frac{\partial}{\partial t}\psi({t, \bf r})={\cal
H}\psi({t, \bf r}),\quad {\bf r}=(x, y,z), \label{eq12}
\ee
where
the Hamiltonian ${\cal H}$ is
 \bb {\cal H} =
\frac{1}{2m}\left(-i\hbar\frac{\partial}{\partial
x}+\frac{eB}{c}y\right)^2
-\frac{\hbar^2}{2m}\frac{\partial^2}{\partial y^2} -
\frac{\hbar^2}{2m}\frac{\partial^2}{\partial z^2}. \label{e12} \ee
Here  $m$; $e$ -- mass and charge of a  particle correspondingly.
The wave function of the particle in field (\ref{e1}) has the
form~\cite{Lan}
 \bb
 \psi_{n p_x p_z}(t, {\bf r})=\frac12 e^{-iE_{n}t/\hbar}e^{i x p_x/\hbar + i z p_z/\hbar
 }U_n(Y),
\label{sol1} \ee where \bb E_{n}=\hbar \omega
\left(n+\frac12\right)+ \frac{{p_z}^2}{2m} \label{e2} \ee is the
energy spectrum of  electron, $\omega=|eB|/mc$,  $p_x$ and $p_z$
are the momenta of the electron in the $x$ and $z$-direction
respectively.

The functions
$$
 U_n(Y) = \frac{1}{(2^n!\pi^{1/2}r_0)^{1/2}}
\exp\left(-\frac{(y-y_0)^2}{2r_0^2}\right)H_n\left(\frac{y-y_0}{r_0}\right),
$$
are expressed through the Hermite polynomials $H_n(z)$, the
integer $n=0, 1, 2, \dots$ indicates the Landau level number,
$r_0=\sqrt{\hbar c/|eB|}\equiv \sqrt{\hbar/m\omega}$  is the
so-called magnetic length (see, for example,~\cite{Ter})  and
$y_0=-cp/eB$.

Now we study a simple solvable model. We consider the motion of an
scalar particle in the case of three dimensions in a single attractive $\delta({\bf r})$
potential and in the presence of a uniform magnetic field. Here
 $\delta({\bf r})$ is the Dirac
delta function. In fact, the equation we need solve is the
following Schr\"odinger equation

\bb
\frac{1}{2m}\left[\left(-i\hbar\frac{\partial}{\partial
x}+\frac{eB}{c}y\right)^2 -\hbar^2\frac{\partial^2}{\partial y^2}
-\hbar^2\frac{\partial^2}{\partial z^2}
-\hbar^2\delta({\bf r})\right]\Psi_{E'}({\bf r})=E'\Psi_{E'}({\bf
r}). \label{e21} \ee
 Solutions of Eq.(\ref{e21}) we can take in the form

\bb \Psi_{E'}({\bf r})= \sum\limits_{n,p_x,p_z}
C_{E'np_x p_z}\psi_{n p_x p_z}({\bf r})\equiv  \sum\limits_{n=0}^{\infty} \int dp_x dp_z
C_{E'np_x p_z}\psi_{np_x p_z}({\bf r}), \label{sum1} \ee where $\psi_{n{p_x}{ p_z}}({\bf
r})$ is the spatial part of the wave functions (\ref{sol1}).

Coefficients $C_{E' n p_x p_z}$ can easily be calculated and then
we obtain the following equation

\bb 1= N \sum\limits_{n=0}^{\infty}\int dp_z\frac{1}{n+A},
\label{ener}
 \ee
 where $ N  $ -  normalized coefficient is
independent on the field and \bb
A=\frac12-\frac{E}{\hbar\omega}+\frac{{p_z}^2}{2 m
{\hbar\omega}}\label{ener1}. \ee Integrating under $p_z$ gives the
equation (\ref{ener}) in the form
\begin{equation}
1=  N \pi \sqrt{2 m \hbar\omega}\,\sum\limits_{n=0}^{\infty}\,\,\,\frac{1}{(n+A)^{1/2}}. \label{cof}
\end{equation}

It is easy to see that Eq.(\ref{cof}) defines implicitly  energy
of a bound localized electron state in the magnetic field. It
should be mentioned that (\ref{cof}) is consistent with analog
result of \cite{Kh2}, where this equation then has been solved
numerically.

However Eq. (\ref{cof}) may be analytically reduced to more simple
form. Indeed the summation with respect to $n$ on the right-hand
side of Eq.(\ref{cof}) can be performed by using the
representation \bb
\frac{1}{(n+A+i\varepsilon)^{1/2}}=\frac{e^{-i\frac{\pi}{4}}}{\sqrt{\pi}}\,\,\,
\int\limits_{0}^{\infty}\frac{e^{i(n+A+i\varepsilon)}}{t^{1/2}}dt.
\label{cof1} \ee As a result Eq.(\ref{cof}) becomes \bb 1 = N_1
\sqrt{\hbar\omega}\,\,\,\frac{e^{-i\frac{3
\pi}{4}}}{2\sqrt{\pi}}\int\limits_{0}^{\infty}\frac{e^{-i\frac{E'
}{\hbar \omega}t}}{t^{1/2}\sin(t/2)}dt, \label{cof2} \ee where
$N_1$- the real constant which is independent on the field.
 As  the required energy
 \bb
 E'= - |E'| ,\label{dop1}
 \ee
must be negative, we can rotate the integration contour
by angle $\pi/2$ in the complex plane of $t$.
Thus we have a real expression
\bb
-1 = N_1 \frac{\sqrt{\hbar\omega}}{2\sqrt{\pi}}\int\limits_{0}^{\infty}\frac{e^{-\frac{E }{\hbar \omega}t}}{t^{1/2}{\rm sinh}(t/2)}dt, \label{cof21}
\ee
where  $E = |E'| \geq 0$.
If we eliminate the magnetic field, the (\ref{cof21}) takes the form

\bb -1 = N_1
\sqrt{\frac{\hbar}{\pi}}\int\limits_{0}^{\infty}\frac{e^{-{E_0 t
/\hbar}}}{t^{3/2}}dt, \label{cof3} \ee where $E_0=|E'_0|$ is the
absolute value of the bound energy of the particle in the
$\delta$-potential without the action of the external field.
Subtracting (\ref{cof3}) from (\ref{cof21}) and removing  of
integrals divergences in the lover limit by the way of standard
regularization procedure we have \bb
      \int\limits_{0}^{\infty}\frac{e^{-E_0 t/\hbar}-e^{-E t/\hbar}}{t^{3/2}}dt= \int\limits_{0}^{\infty}\frac{e^{-E t/\hbar}}{t^{3/2}}\left( \frac{a_1 t}{{\rm {sinh}}{(a_1 t)}}-1\right)dt, \label{cof4}
\ee
where $a_1=\frac{\omega}{2 } $. From (\ref{cof4})  it is easily to obtain

\bb
\sqrt{E}-\sqrt{E_0}=\frac{\sqrt{E}}{2\sqrt{\pi}}\int\limits_{0}^{\infty}\frac{e^{-
x}}{x^{3/2}}\left( \frac{a x}{{\rm {sinh}}{(a x)}}-1\right)dx,
\label{cof5} \ee where $a=\frac{\hbar\omega}{2 E } $, that is
consistent with analog equation obtained by the well-known method
using  boundary conditions of wave functions in the model of
$\delta$-potential~\cite{Dem.2},\cite{Pop},\cite{RoKr3}.

In the weak field limit $\hbar\omega \ll 2 E_0$ developing the
expansion of the integrand function in (\ref{cof5})  we can obtain
\bb
       E = E_0 \left(1-\frac{1}{48}\frac{{\hbar}^2{\omega}^2}{{E_0}^2} + \frac{1}{576}\frac{{\hbar}^4{\omega}^4}{{E_0}^4}\right). \label{cof51}
\ee Note that the square term  in (\ref{cof51}) is coincided with
analog result of \cite{Dem.2}.

In order to consider the case of a strong field  $\hbar\omega > 2
E_0$ we produce the right-hand side of (\ref{cof5}) to the
analytic form \bb
        -\frac{1}{\sqrt{a_0}}=\frac{1}{\sqrt{2}}{\rm{Zeta}}\left[\frac{1}{2},\frac{1}{2}+\frac{1}{2 a}\right], \label{cof6}
\ee where $ a_0 = \frac{\hbar\omega}{2 E_0}$ and
$\rm{Zeta[\nu,p]}$ is a generalized Riemann Zeta-function. The
range of the validity (\ref{cof6}) be found some wider as was
supposed initially. In fact for derivation of (\ref{cof21}) we
assume that $E' \leq 0$, however from (\ref{cof6}) it should be
that argument of Zeta-function can be continuously reached the
values
$$
1/2 + 1/2a > 0 .
$$
This condition gives the limitation of the
required bound energy spectrum \bb
   E' < \frac{\hbar\omega}{2}. \label{cof61}
\ee
The physical meaning of this condition consists of restriction to the continuous spectrum of the scalar
particle in the magnetic field by the value (\ref{cof61}).
  Note that
after change of variables in (\ref{cof6}) that is coincided with the basic equation of work \cite{Dem.2}
where the case of scalar particles in the magnetic field was considered and analog conclusion about
limitation of continuous spectrum has been done.

Expansion of $\rm {Zeta[\nu,p]}$ in the limit $ p \ll 1$ gives the result
\bb
   \rm{ Zeta}[1/2,p]=\frac{1}{p^{1/2}}+\rm{Zeta}[1/2]-\frac{1}{2}\rm{Zeta}[3/2]p +\frac{3}{8}\rm{Zeta}[5/2]p^2 + 0[p]^3. \label{cof7}
\ee
 After substitution of (\ref{cof7}) to (\ref{cof6}) we can
obtain the equation that explicitly determines the bound state
energy in the strong field limit

\bb
   {E'}=  \hbar\omega\left( 0.205-0.452\sqrt{\frac{E_0}{\hbar\omega}}
   -0.367\, \frac{E_0}{\hbar\omega}\right).\label{cof8}
\ee It should be emphasized that in  the super strong  magnetic
fields  expansion (\ref{cof8}) gives the upper limit of binding
energy of the scalar particle
$$
           E' =  0.205\, \hbar\omega\,,
           $$
which is no contradicted to the condition (\ref{cof61}).
Furthermore one can see that this limited value is not dependent
on the particle energy in the absence of the field, but it is
completely  determined by the magnetic field intensity.

It is of interest to compare the obtained results with the case of two-dimensional model.
The analog of the Eq.(\ref{cof}) in the two-dimensional case takes the form

\bb
   1=  \frac{1}{8\pi}\int\limits_{0}^{\infty}\frac{e^{-E t/\hbar\omega}}{{\rm sinh}(t/2)}dt, \label{cof9}
\ee
 what is coinciding with the corresponding result of the work
\cite{Kh1}. However, unlike treatment \cite{Kh1} the
regularization procedure  we carry out in another way. Thus as
early (see (\ref{cof4})) we remove  the magnetic field and obtain
\bb 1= \frac{1}{4\pi}\int\limits_{0}^{\infty}\frac{e^{-E_0
t}}{t}dt. \label{cof10} \ee Carrying out a simple calculation
which is similar to the described regularization procedure for the
three-dimensional case we can write \bb
     \ln\frac{E}{E_0}=\int\limits_{0}^{\infty}\frac{e^{-x}}{x}\left( \frac{a x}{{\rm sinh}(a x)}-1 \right) dx, \label{cf1}
\ee
where as before $a=\hbar\omega/(2 E)$.
In the weak field limit  from (\ref{cf1}) we have
\bb
      E=E_0 \left(1- \frac{\hbar^2\omega^2}{24 {E_0}^2}\right).\label{cf2}
\ee

For the successive consideration of the range $\hbar\omega > 2
E_0$ at first the integral in the right-hand side of the
Eq.(\ref{cf1}) should be calculated analytically

\bb -\ln\left( \frac{E}{E_0} \right)= \ln(2 a)+
\Psi\left(\frac{1+a}{2a} \right), \ee
 where $\Psi(x)$ is a logarithmic derivative of Euler Gamma function.
 Then we have the basic equation in the two-dimensional model
 \bb
- \ln\left( 2 a_0 \right)=\Psi\left( \frac{1}{2}+\frac{1}{2 a} \right), \label{cf3}
 \ee
where $a_0=\hbar\omega/(2 E_0).$

 In the strong field limit, after evaluation of  $\Psi(p)$ function
\bb
\Psi (p) =-\frac{1}{p} -C +\frac{\pi^2}{6} p + \frac{1}{2}{\rm PolyGamma}[2,1] p^2 +\frac{\pi^4}{90} p^3 +0[p]^4, \label{dop2}
\ee
where
$$
{\rm PolyGamma}[n,z] = \Psi^{(n)}(z) = \frac{d^n \Psi(z)}{d z^n},
$$
 (\ref{cf3}) can be written in the form
 \bb
 \ln{\frac{\hbar\omega}{E_0}} - \frac{1}{\frac{1}{2}-\frac{E'}{\hbar\omega}}- C +\frac{\pi^2}{6}
 \left( \frac{1}{2}-\frac{E'}{\hbar\omega}\right)= 0, \label{cf41}
 \ee
 where  $C = 0.577...$ is Euler constant.

The solution of Eq.(\ref{cf41}) is explicitly determining the
bound state energy can be written as \bb
\frac{E'}{\hbar\omega}=\frac{1}{2}- \frac{6 (C
-\ln(\hbar\omega/E_0))+\sqrt{24 \pi^2 +36(C
-\ln(\hbar\omega/E_0))^2}}{2 \pi^2}.\label{cf5} \ee
 In the limit
$\ln(\hbar\omega/E_0)\gg 1 $  from (\ref{cf5})  we have \bb
\frac{E'}{\hbar\omega}
=\frac{1}{2}-\frac{1}{\ln(\hbar\omega/E_0)}- \frac{C
}{\ln^2(\hbar\omega/E_0)}+\frac{\left( \frac{\pi^2}{6}-C
\right)}{\ln^3(\hbar\omega/E_0)} + 0[\ln(\hbar\omega/E_0)]^4.
\label{dop3} \ee
 Considering the properties of $\Psi(z)$ we see
again that expansion (\ref{dop3}) is correct under the bound
energy $ E' =  \le \hbar\omega /2$. Furthermore this limited value
as before is not dependent on the particle energy in the absence
of the field.  However, there is the essential difference from the
three-dimensional case. Indeed, the  upper limit of the shifted
binding energy level in consider model for the range of supper
strong magnetic fields (when large not only ratio
$\hbar\omega/E_0$, but also $\ln(\hbar\omega/E_0)
>>1)$  has tendency directly to the
boundary of the continuous spectrum.

\section{An electron in an attractive potential in the presence
of a uniform magnetic field}

It is vary important that with the help of the present approach we
can study the spin-effects in the magnetic fields by the same way.
The case of spin-$1/2$ particle can be calculated on the basis of
exact solutions of Pauli equation. The Pauli equation in the field
(\ref{e1}) has the form

\bb i\hbar\frac{\partial}{\partial t}\psi({t, \bf r})={\cal
H}\psi({t, \bf r}),\quad {\bf r}=(x, y,z), \label{eq12}
\ee
where
the Hamiltonian ${\cal H}$ is \bb {\cal H} =
\frac{1}{2m}\left(-i\hbar\frac{\partial}{\partial
x}+\frac{eB}{c}y\right)^2
-\frac{\hbar^2}{2m}\frac{\partial^2}{\partial y^2} -
\frac{\hbar^2}{2m}\frac{\partial^2}{\partial z^2}+ \mu\sigma_3B.
\label{e12} \ee Here  $\mu=|e|\hbar/2m c$ is the Bohr magneton, $m$
is the mass of a  electron and
$$
\sigma_3=\left(\begin{array}{cc}
1 & 0\\
0 &-1\\
\end{array}\right)
$$
is the  $z$-component of Pauli matrixes. The last term in
(\ref{e12}) describes the interaction of the spin magnetic moment
of the electron with the magnetic field. Electron wave function in
field (\ref{e1}) has the form \bb
 \psi_{n p_x p_z s}(t, {\bf r})= \frac{1}{2}\psi_{n p_x p_z}(t, {\bf r})
\left( \begin{array}{c}
1+s\\
1-s
\end{array}\right),
\label{sol2}
\ee
 where $\psi_{n p_x p_z}(t, {\bf r})$ is the solution of the Schr\"{o}dinger equation in
 the field (\ref{e1}) (see(\ref{sol1})),
 \bb E_{ns}=\hbar \omega
\left(n+\frac12\right)+ \frac{{p_z}^2}{2m}+ s\hbar
\omega\frac{1}{2} \label{e2} \ee is the energy spectrum of
electron, $\omega=|eB|/mc$, $s=\pm 1$ is conserving spin quantum
number, $p_x$ and $p_z$ are the momenta of the electron in the $x$
and $z$-direction respectively.

It is vary important that ground electron state in
the magnetic field essentially different from the analog state of spin-$0$ particles.
At the same time the boundaries of continuous spectra
for spinor particles will be differ from scalar one. For example, if continuous spectrum
of scalar particle begin from value $E' \geq \hbar \omega/2$,  for an
electron with spin directed along the magnetic field analog condition is
 $E'\geq \hbar \omega$.
For particle with spin opposite to the magnetic field
strength continuous spectrum is beginning from values $E' \geq 0$.

 Thus taking into account an interaction
of the electron spin magnetic moment  with the magnetic field the
equation for energy in the three-dimensional case can be written
in the form

 \bb
 \sqrt{E}-\sqrt{E_0}= \frac{\sqrt{E}}{2\sqrt{\pi}}\int\limits_{0}^{\infty}\frac{e^{-t }}
 {t^{3/2}}\left(\frac{a t e^{- s a t}}{{\rm sinh}(a t)}-1\right)dt, \label{cf6}
\ee
were $s=\pm 1$ corresponds to spin orientations
 along and
opposite to the direction of the magnetic field respectively.
Expend the integral in (\ref{cf6}) in the limit $a << 1$ we have
the equation \bb \sqrt{E} -\sqrt{E_0}=
-\frac{s\hbar\omega}{4\sqrt{E}}+\frac{\sqrt{E}}{12}\left(\frac{\hbar\omega}{2
E_0}\right)^2.\label{cf62} \ee The solution of Eq.(\ref{cf62}) has
the form \bb
           \sqrt{\frac{E}{E_0}}=\frac{2\left( -12 E_0 -\sqrt{3}\sqrt{48 {E_0}^4+
           s(\hbar^3\omega^3 E_0 -48 \hbar\omega {E_0}^3}\right)}{\hbar^2\omega^2 -48 {E_0}^2}.\label{cf63}
\ee
The expansion  (\ref{cf63}) in
the weak field limit can be written as follows:

\bb
 \frac{E}{E_0}=1 - s\frac{\hbar\omega}{2 E_0}-\frac{1}{48} \frac{\hbar^2\omega^2}{E_0^2}. \label{cf61}
\ee From (\ref{cf61}) it is easy to see that the level of energy
$E'_0=-|E_0|$, existing in $\delta$-potential without any
perturbation, for the case $s=1$  under the action of the magnetic
field increases  at  $ \frac{\hbar\omega}{2 E_0}$,  and for $s=-1$
it falls at $\frac{\hbar\omega}{2 E_0}$ in  terms of the negative
energetic  scale. However the depth of energetic levels with
respect to the boundaries of the continuous spectra  for this two
cases are equal among themselves and equal to the case of the
spin-$0$ particles.

Integrating of the right-hand side Eq.(\ref{cf6}) we can obtain the equation in the analytical form
\bb
   -\frac{1}{\sqrt{a_0}}=\frac{1}{\sqrt{2}}\rm {Zeta}\left[\frac{1}{2},\frac{1}{2}+
   \frac{s}{2}+\frac{1}{2 a}\right]. \label{cf7}
\ee
In the strong field limit $ \hbar\omega > E_0$  one can write
the generalized Riemann's Zeta-function in Eq.(\ref{cf7}) as
follows \bb
      \frac{1}{\sqrt{2}}\rm {Zeta}\left[\frac{1}{2},\frac{1}{2}+\frac{s}{2}+\frac{1}{2 a}\right]= \frac{1}{\sqrt{2}}\frac{1}{\sqrt{\frac{1+s}{2}+\frac{E}{\hbar\omega}}}+\frac{\rm {Zeta[1/2]}}{\sqrt{2}}-\frac{\rm{Zeta[3/2]}}{2\sqrt{2}}\left(\frac{1+s}{2}+\frac{E}{\hbar\omega}\right).
\ee
Finally, the Eq.(\ref{cf7}) can be written in the following way
\bb
    \frac{1}{\sqrt{2}} +\left( \sqrt{\frac{2 E_0}{\hbar\omega}}+\frac{\rm{Zeta}[1/2]}{\sqrt{2}}\right)x -
    \frac{\rm{Zeta}[3/2]}{2 \sqrt{2}} x^3 = 0, \label{cf8}
\ee where $$x = \sqrt{\frac{1+s}{2}+\frac{E}{\hbar\omega}}.$$
Solutions of Eq.(\ref{cf8}) for different spin values
$ s =0, \,
+1 ,\, -1$
 are represented as

  \bb
      E' =  \hbar \omega \left( 0.205 + \frac{s}{2} -0.452 \sqrt{\frac{E_0}{\hbar\omega}}-
      0.367 \frac{E_0}{\hbar\omega}  \right). \label{cf81}
\ee
 Figure 1. presents a comparison of the graphic solutions of
Eq.(\ref{cf7}) for different values of spin  under $E_0 = 1$ and
$\hbar\omega = 100 E_0$. It is evidently, that approximate
solutions (\ref{cf81}) give results which are near the explicit
points of intersections of left-hand and right-hand sides  of the
Eq.(\ref{cf7}). It should be emphasized that in strong field limit
the dependence of energetic level shifts from the spin is not
disappeared. However in all cases displacements of binding energy
levels as in the weak field limit  are at the equal distances with
respect to the boundaries of the continuous spectra.

Let us now consider the two-dimensional case with regard for  spin
interactions. According to present approach  we can write

\bb \ln
\left(\frac{E}{E_0}\right)=\int\limits_{0}^{\infty}\frac{e^{-x}}{x}
\left( \frac{a x e^{- s a x}}{{\rm sinh}(a x)} -1\right) dx,
\label{cf91} \ee where the direction of spin particles as before
is characterized by $s=\pm 1$. In the range of weak fields from
Eq.(\ref{cf91}) we have
 \bb
                  \frac{E}{E_0}= 1-\frac{s}{2}\frac{\hbar \omega}{E_0}-
                  \frac{1}{24}\left( \frac{\hbar \omega}{E_0}\right)^2 .\label{cf9}
\ee
 For consideration of opposite  limit  it is
clear we must calculate the integrals in the above equation in the
analytical form
\bb \int\limits_{0}^{\infty}\frac{e^{-x}}{x}
\left( \frac{a x e^{- s a x}}{{\rm sinh}(a x)} -1\right) dx =
-2a\frac{(1+ s)}{2}-\ln(2a)-\Psi\left( \frac{1}{2 a }\right).
\label{cf10}
 \ee
 Then equations for energy displacements for
$\hbar\omega > E_0 $ will be written as \bb
          \ln\left( \frac{E}{E_0} \right)= 2a\frac{(1-s)}{2}-\ln(2a)+ C -\frac{\pi^2}{12 a}. \label{cf11}
\ee
 For the case $s= -1$ in the strong field limit ($ \ln\frac{\hbar\omega}{E_0} >> 1 $) from Eq.(\ref{cf91}) at once may be
obtained
 \bb
    E' = -\hbar\omega\left(\frac{1}{\ln\frac{\hbar\omega}{E_0}}+\frac{C}{\ln\frac{\hbar\omega}{E_0}}\right).\label{cf101}
\ee

For the opposite spin orientation ($ s=1$) in Eq.(\ref{cf10}) we
at first must use the recurrent relation for $\Psi (p)$ function

\bb
        \frac{1}{x} +  \Psi (x) = \Psi (1+ x).
\ee
After that using the asymptotic expansion for $ \Psi (p)$ (see (\ref{dop2}))we also have
\bb
        E' = \hbar\omega\left(1 - \frac{1}{\ln\frac{\hbar\omega}{E_0}}-\frac{C}{(\ln\frac{\hbar\omega}{E_0})^2}\right).\label{cf101}
\ee

It is ease to see that  dependence from spin parameters can be
interpreted  as before in three-dimensional model. The main
difference from this case that in the supper strong magnetic field
we have convergence of  the considered binding energy levels to
the boundaries of continuous spectra.

\section{Discussion and conclusions}

We have shown that the effect of a magnetic field on localized
electron states leads to the number of equations for the binding
energy of spin-$0$ and spin-$1/2$ particles. Thus, in the weak
field limit energy displacements scalar and spinor  particles  in
the case of the three dimensional model can be described by the
following expressions \bb
  \left. \begin{array}{rcl}
 s=0\,\,\,\,\,\, \frac{\hbar\omega}{2 E_0}&+&\frac{E}{E_0}\\
 s=1\,\,\,\,\,\,\,\,\frac{\hbar\omega}{E_0} &+& \frac{E}{E_0}\\
 s=-1\,\,\,\,\,\,\,\,\,\,\,\,\,\, &{}& \frac{E}{E_0}\\
  \end{array}\right \} = 1 + \frac{\hbar\omega}{2 E_0} - \frac{\hbar^2\omega^2}{48 {E_0}^2}. \label{car1}
\ee

For the two-dimensional case in this limit we can write

\bb
  \left. \begin{array}{rcl}
 s=0\,\,\,\,\,\, \frac{\hbar\omega}{2 E_0}&+&\frac{E}{E_0}\\
 s=1\,\,\,\,\,\,\,\,\frac{\hbar\omega}{E_0} &+& \frac{E}{E_0}\\
 s=-1\,\,\,\,\,\,\,\,\,\,\,\,\,\, &{}& \frac{E}{E_0}\\
  \end{array}\right \} = 1 + \frac{\hbar\omega}{2 E_0} - \frac{\hbar^2\omega^2}{24 {E_0}^2}. \label{car1}
\ee

In the strong field the dependence on spin is not disappeared. In
three-dimensional model energetic levels approach to specific
values of spectra, which determine the binding states. For
different values of  particle spin the displacements levels of
binding energy are at the equal distances with respect to
boundaries of   the continuous spectra. In this connection for the
tree-dimensional case it can be represented in the form

\bb
  \left. \begin{array}{rcl}
s=0\,\,\,\,\,\,\,\,\, \frac{\hbar\omega}{2}&-&E'\\
s=1\,\,\,\,\,\,\,\,\, \hbar\omega    & - &E'     \\
s=-1\,\,\, \,\,\,\,\,\,\,\,\,\,\,\,\,\,       &{-} & E'    \\
  \end{array}\right \} =  \hbar\omega\left( 0.295 +  0.452 \sqrt{\frac{E_0}{\hbar\omega}}+
  0.367 \frac{E_0}{\hbar\omega}\right). \label{car2}
\ee Hence for cases spinless particles $s=0$ and for the electron
with $s = 1$ in strong fields we have positive values of the
binding energy levels whereas for electrons with $s=-1$ it remains
negative.

It is ease to see that  dependence on spin parameters in
two-dimensional model can be  written in the analogical form

 \bb
  \left. \begin{array}{rcl}
s=0\,\,\,\,\,\,\,\,\, \frac{\hbar\omega}{2}&-&E'\\

s=1\,\,\,\,\,\,\,\,\, \hbar\omega    & - &E'     \\

s=-1\,\,\, \,\,\,\,\,\,\,\,\,\,\,\,\,\,       &{-} & E'    \\
  \end{array}\right \} =  \frac{\hbar\omega}{\ln\frac{\hbar\omega}{E_0}}+
  \frac{C\,\, \hbar\omega}{(\ln\frac{\hbar\omega}{E_0})^2} \label{car3}
\ee

From Eq.(\ref{car3}) one can see that in this limit the energy
levels also is not  dependent on the energy of the particle in the
absence of the field. The distinctive feature of this case is the
binding energy levels for the range of supper strong magnetic
fields when $\ln(\hbar\omega/E_0)
>>1$ are directly approaching to the
boundaries of the continuous spectra for all considering spin
values.

It is shown  that shifts of energy levels of a polarized electron
arising under the action of a weak magnetic field as in the three
or two-dimensional models for different values of the particle
spin go on by the similar way. We have the line displacements as
the themselves levels for $s=1$ and $s=-1$ and analogical shifts
of the boundaries of continuous spectra for $s=1$. The same
picture we have and in the case of spinless particle where the
line shift of the continuous spectrum boundary takes place. It is
clear that in case of the weak intensity a magnetic field indeed
plays a stabilizing role in considering systems, because the depth
of the binding energy levels is increased under the field action
independently on the particle spin. However in the strong field
limit our results showered nonlinear dependence on the field
intensity of the level displacements. Nevertheless in this limit
the conditions are determined the boundaries of continuous spectra
as before have linear dependence on the field. Then in supper
strong magnetic fields one may fined that binding energy levels
approaching to the boundaries of continuous spectra. The existing
distinctions can be formulated in the following way. In the model
of (3+1) dimension there is limiting positions of energy levels,
which are at fixed distances from boundaries  of continuous
spectra. In case of two-dimensional model energy levels in a super
strong magnetic field asymptotic aspire to boundaries  of a
continuous spectra. But in both cases we have increasing of
instability of systems in strong magnetic fields. So this
conclusion is disproved the common opinion about a permanently
stabilizing role of a magnetic field in ionization processes.

\end{document}